\documentclass{ws-procs9x6}
\usepackage{graphicx}

\newcommand{\gap}{\mathrel{ \rlap{\raise.5ex\hbox{$>$}}
                    {\lower.5ex\hbox{$\sim$}}  } }
\newcommand{\lap}{\mathrel{ \rlap{\raise.5ex\hbox{$<$}}
	            {\lower.5ex\hbox{$\sim$}}  } }
\newcommand{\ob}{$\Omega_b$}
\newcommand{\obh}{$\Omega_bh^2$}

\newcommand{\deu}{$D$}
\newcommand{\tro}{$^3He$}
\newcommand{\qua}{$^4He$}
\newcommand{\six}{$^{6}Li$}
\newcommand{\sep}{$^{7}Li$}

\newcommand{\zaa}{{\it Astron. Astrophys.}}

\newcommand{\zapj}{{\it Astrophys. J.}}
\newcommand{\znp}{{\it Nucl.~Phys.}}
\newcommand{\zzp}{{\it Z.~Phys.}}
\newcommand{\zpl}{{\it Phys.~Lett.}}
\newcommand{\zpr}{{\it Phys.~Rev.}}

\newcommand{\zadndt}{{\it Atomic Data and Nuclear Data Tables}}

\newcommand{\znas}{{\it New~Astronomy}}
\newcommand{\znat}{{\it Nature}}

\newcommand{\zapjs}{{\it Astrophys.~J.~S.}}

\begin{document}

\title{Updated Big--Bang Nucleosynthesis compared to WMAP results}

\author{Alain~Coc}

\address{CSNSM, CNRS/IN2P3/UPS, B\^at.~104, 91405 Orsay Campus, France}

\author{Elisabeth~Vangioni-Flam}

\address{IAP/CNRS, 98$^{bis}$ Bd. Arago, 75014 Paris France}

\author{Pierre~Descouvemont}

\address{Physique Nucl\'{e}aire Th\'eorique et Physique Math\'ematique, 
CP229,\\
Universit\'{e} Libre de Bruxelles, B-1050 Brussels, Belgium}

\author{Abderrahim~Adahchour\footnote{Permanent address: LPHEA, 
FSSM, Universit\'e Caddi Ayyad, Marrakech, Morocco}}

\address{Physique Nucl\'{e}aire Th\'eorique et Physique 
Math\'ematique, CP229,\\
Universit\'{e} Libre de Bruxelles, B-1050 Brussels, Belgium}

\author{Carmen~Angulo}

\address{Centre de Recherche du Cyclotron,  
  UcL, Chemin du Cyclotron 2,\\ 
  B-1348 Louvain--La--Neuve, Belgium}

\maketitle

\abstracts{
From the observations of the anisotropies of the Cosmic Microwave 
Background (CMB) radiation, the WMAP satellite has provided a determination 
of the baryonic density of the Universe, \obh,  with an unprecedented 
precision. 
This imposes a 
careful reanalysis of the standard Big--Bang Nucleosynthesis (SBBN) 
calculations. We have updated our previous calculations using  
thermonuclear reaction rates provided by a new analysis of 
experimental nuclear data constrained by $R$-matrix theory. 
Combining these BBN
results with the \obh\ value from WMAP, we deduce the light element 
(\qua, \deu, \tro\ and \sep) primordial abundances and compare them
with spectroscopic observations. There is a very good agreement   
with deuterium observed in cosmological clouds,
which strengthens the confidence on the estimated baryonic
density of the Universe.
However, there is an important discrepancy between the deduced
\sep\ abundance and the one observed in halo stars of our Galaxy, 
supposed, until now, to represent the primordial abundance of this 
isotope. 
The origin of this discrepancy, observational, nuclear or more 
fundamental remains to be clarified.
The possible role of the up to now neglected 
$^7$Be(d,p)2$\alpha$ and $^7$Be(d,$\alpha$)$^5$Li reactions is considered.
}


\section{Introduction}

Big--Bang nucleosynthesis used to be the only method to determine the baryonic
content of the Universe. However, recently other methods have emerged. In 
particular the analysis of the anisotropies of the cosmic microwave 
background radiation has provided \obh\ values with ever increasing 
precision.   
(As usual, \ob\ is the ratio of the baryonic density over the critical density
and  $h$ the Hubble constant in units
of 100~km$\cdot$s$^{-1}$$\cdot$Mpc$^{-1}$.) 
The baryonic density provided by WMAP\cite{WMAP}, \obh\ = 0.0224$\pm$0.0009, 
has indeed dramatically increased the precision on this crucial cosmological
parameter with respect to earlier experiments: BOOMERANG, CBI, DASI, 
MAXIMA, VSA and ARCHEOPS. 
It is thus important to improve the precision on SBBN calculations. 
Within the standard model of BBN, the 
only remaining free parameter is the baryon over photon ratio $\eta$ 
directly related to \obh\ [\obh=3.6519$\times10^7\;\eta$].
Accordingly, the main source of uncertainties comes from the nuclear 
reaction rates.
In this paper we use the results of a new analysis\cite{Des03,Coc03} of 
nuclear data providing 
improved reaction rates which reduces those uncertainties.

\section{Nuclear reaction rates}

In a previous paper\cite{Coc02} we already used a Monte--Carlo 
technique, 
to calculate the uncertainties on the light element yields (\qua, \deu, \tro\ and 
\sep) related to nuclear reactions.
The results were compared to observations that are thought to be
representative of the corresponding primordial abundances. 
We used reaction rates from the NACRE 
compilation of charged particles reaction rates\cite{NACRE}  
completed by other sources\cite{Smi93,Bru99,Che99} as NACRE
did not include all of the 12 important reactions of SBBN.    
One of the main innovative features of NACRE with respect to former
compilations\cite{CF88} is that  uncertainties are analyzed in detail and
realistic lower and upper bounds for the rates are provided.
However, since it is a general compilation for multiple applications, 
coping with a broad range of nuclear configurations, 
these bounds had not always been evaluated through a rigorous 
statistical methodology.
Hence, we assumed a simple uniform distribution between these 
bounds for the Monte--Carlo calculations. 
Other works \cite{NB00,Cyb01} have 
given better defined statistical limits for the reaction rates of 
interest for SBBN. 
In these works, the astrophysical $S$--factors (see definition in 
Ref.~\cite{Coc02}) were either fitted with spline functions\cite{NB00}  
or with NACRE $S$--factor fits and data but using a different 
normalization\cite{Cyb01}.
In this work, we use a new compilation\cite{Des03} specifically 
dedicated to SBBN reaction rates using for the first time in this context 
nuclear theory to constrain the $S$--factor energy dependences and provide 
statistical limits.
The goal of the $R$-matrix method\cite{LT58} is to parametrize some 
experimentally known quantities, such as cross sections or phase shifts, 
with a small number of parameters, which are then used to interpolate 
the cross section within astrophysical energies. 
The $R$-matrix theory has been used for many decades in the nuclear physics
community (see e.g. Ref.~\cite{Bar91,Ang98} for a recent application 
to a nuclear astrophysics problem) but this is the first time that it is 
applied to SBBN reactions.
This method can be used for both resonant and non-resonant contributions 
to the cross section. (See Ref.~\cite{Des03} and references therein for 
details of the method.) 
The $R$-matrix framework assumes that the space is divided into two 
regions: the internal region (with radius $a$), where nuclear forces 
are important, and the external region, where the interaction between the 
nuclei is governed by the Coulomb force only. 
The physics of the internal region is parameterized by a number $N$ of poles,
which are characterized by energy $E_{\lambda}$ and reduced width
$\tilde{\gamma}_{\lambda}$.
Improvements of current work on Big Bang nucleosynthesis essentially 
concerns a more precise evaluation of uncertainties on the reaction 
rates. Here, we address this problem by using standard statistical 
methods \cite{PDG96}.
This represents a significant improvement with respect to NACRE \cite{NACRE}, 
where uncertainties are evaluated with a simple prescription.  
The $R$-matrix approach depends on a number of parameters, some of them 
being fitted, whereas others are constrained by well determined data, 
such as energies or widths of resonances.
As usual, the adopted parameter set is obtained from the minimal $\chi^2$ 
value. 
The uncertainties on the parameters are evaluated as explained in 
Ref.\cite{PDG96}. The range of acceptable $p_i$ values is such that
$\chi^2(p_i) \leq \chi^2(p_i^{min}) + \Delta \chi^2$,
where $p^{min}_{i}$ is the optimal parameter set. 
In this equation, $\Delta \chi^2$ is obtained from 
$P(\nu/2, \Delta \chi^2/2)=1-p$,
where $\nu$ is the number of free parameters $p_i$,
$P(a,x)$ is the Incomplete Gamma function, and $p$ is the confidence 
limit ($p=0.683$ for the $1\sigma$
confidence level)\cite{PDG96}.
This range is scanned for all parameters, and the limits on the
cross sections are then estimated at each energy. 
As it is well known, several reactions involved in nuclear astrophysics 
present different data sets which are not compatible
with each other. An example is the $^3$He($\alpha,\gamma)^7$Be
reaction where data with different normalizations are available.
In such a case, a special procedure is used\cite{Des03}.

This new compilation\cite{Des03} provides 
1--$\sigma$ statistical limits for each of the 10 rates: 
$^2$H(p,$\gamma)^3$He, $^2$H(d,n)$^3$He, $^2$H(d,p)$^3$H, $^3$H(d,n)$^4$He
$^3$H($\alpha,\gamma)^7$Li, $^3$He(n,p)$^3$H, $^3$He(d,p)$^4$He, 
$^3$He($\alpha,\gamma)^7$Be, $^7$Li(p,$\alpha)^4$He and $^7$Be(n,p)$^7$Li.
The two remaining reactions of importance, n$\leftrightarrow$p and 
$^1$H(n,$\gamma)^2$H come from theory and are unchanged with respect to our 
previous work\cite{Coc02}.

\section{SBBN calculations}

\begin{figure}[h]
\vskip -4cm
\includegraphics[width=10.7cm]{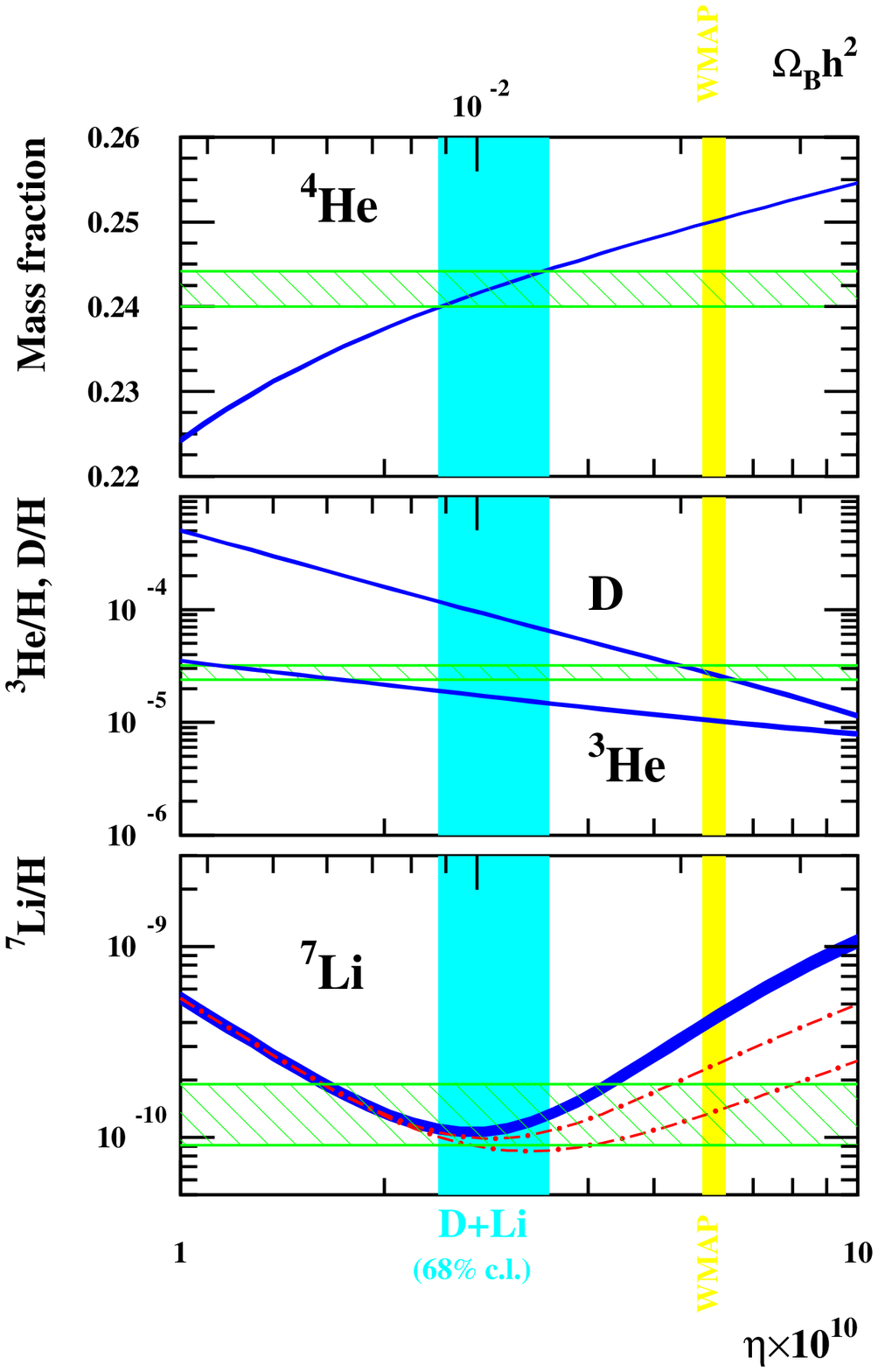}
\caption{Abundances of \qua\ (mass fraction), \deu, \tro\ and \sep\ (by
number relative to H) as a function of the baryon over photon ratio 
$\eta$ or \obh. Limits (1-$\sigma$) are obtained from Monte Carlo 
calculations. 
Horizontal lines represent primordial \qua, \deu\ and \sep\ abundances deduced
from observational data (see text). 
The vertical stripes represent the (68\% c.l.) 
\obh\ limits provided by WMAP\protect\cite{WMAP} or deduced from
\sep\ and \qua\ observations and SBBN calculations. 
For the dash-dotted lines in the bottom panel: see text.}
\label{f:bbn}
\end{figure}

We performed Monte-Carlo calculations using Gaussian 
distributions with 
parameters provided by the new compilation and calculated the  
\qua, \deu, \tro\ and \sep\ yield range as a function of $\eta$, fully
consistent with our previous analysis\cite{Coc02}.   
The differences with Ref.~\cite{Cyb01} on the \sep\ yield is probably 
due to their different normalization procedure of the NACRE $S$--factors.
Figure~\ref{f:bbn} displays the resulting abundance limits (1-$\sigma$) 
[it was 2-$\sigma$ in Fig.4 of Ref.~\cite{Coc02}] from SBBN calculations 
compared to primordial ones inferred from observations.
Using these results and the WMAP \obh\ range (quoted WMAP+SBBN in the 
following), it is now possible  to infer the primordial \qua, \deu, 
\tro\ and \sep\ abundances.

We obtain  (WMAP+SBBN) a deuterium primordial abundance of
D/H = $(2.60^{+0.19}_{-0.17})\times10^{-5}$ 
[ratio of D and H abundances by number of atoms]
which is in perfect agreement with the average value 
$(2.78^{+0.44}_{-0.38})\times10^{-5}$ of D/H observations in 
cosmological clouds\cite{Kir03}. 
These clouds at high redshift on the line of sight of 
distant quasars are expected to be representative of primordial \deu\ 
abundances.
The exact convergence between these two independent methods is claimed 
to reinforce the confidence in the deduced \obh\ value.

\clearpage

The other WMAP+SBBN deduced primordial abundances 
are $Y_P$ = 0.2457$\pm0.0004$ for the \qua\ mass fraction, 
$^3$He/H = $(1.04\pm0.04)\times10^{-5}$ and $^7$Li/H = 
$(4.15^{+0.49}_{-0.45})\times10^{-10}$.
We leave 
aside \tro\ whose primordial abundance cannot be reliably determined
because of its uncertain rate of stellar production and 
destruction\cite{EVF03}.

The \qua\ primordial abundance, $Y_p$ (mass fraction), is derived from 
observations of metal--poor extragalactic, ionized hydrogen (H~II) regions.
Recent evaluations gave a relatively narrow ranges of abundances:  
$Y_p$ = 0.2452$\pm$0.0015 (Izotov et al.\cite{Izo99}), 
0.2391$\pm$0.0020 (Luridiana et al.\cite{Lur03}).
However, recent observations by Izotov and Thuan\cite{Izo04} on a large
sample of 82 H~II regions in 76 blue compact galaxies have lead to the 
value of $Y_p$ = 0.2421$\pm$0.0021 that we adopt here.
With this range, WMAP and SBBN results are hardly compatible.
Nevertheless, as systematic uncertainties may prevail 
due to observational difficulties and complex physics\cite{Fie98}
\qua\ alone is unsufficient to draw a conclusion.

The \sep\ abundance measured in halo stars of the Galaxy is considered up 
to now as representative of the primordial abundance as it display a 
plateau\cite{Spi82} as a function of metallicity (see definition in 
Ref.~\cite{Coc02}). Recent observations\cite{Rya00} have lead to (95\% c.l.)  
Li/H = $(1.23^{+0.68}_{-0.32})\times10^{-10}$. 
These authors have
extensively studied and quantified the various sources of uncertainty : 
extrapolation, stellar depletion and stellar atmosphere parameters.
This Li/H value, based on a much larger number of observations than 
the D/H one
was considered\cite{Coc02} as the most reliable constraint on SBBN and 
hence on \obh. 
However, it is a factor of 3.4 lower than the WMAP+SBBN value. 
Even when considering the corresponding uncertainties, the two Li/H 
values differ drastistically.
This confirms our\cite{Coc02} and other\cite{Cyb01,Cyb03} previous 
conclusions that the \obh\ range deduced from SBBN of \sep\ are only 
marginally 
compatible with those from the CMB observations available by this time
(BOOMERANG, CBI, DASI and MAXIMA experiments).
It is strange that the major discrepancy affects \sep\ since 
it could a priori lead to a more reliable primordial value than deuterium, 
because of much higher observational statistics and an easier extrapolation 
to primordial values.

\begin{figure}[htb]
\includegraphics[width=12cm]{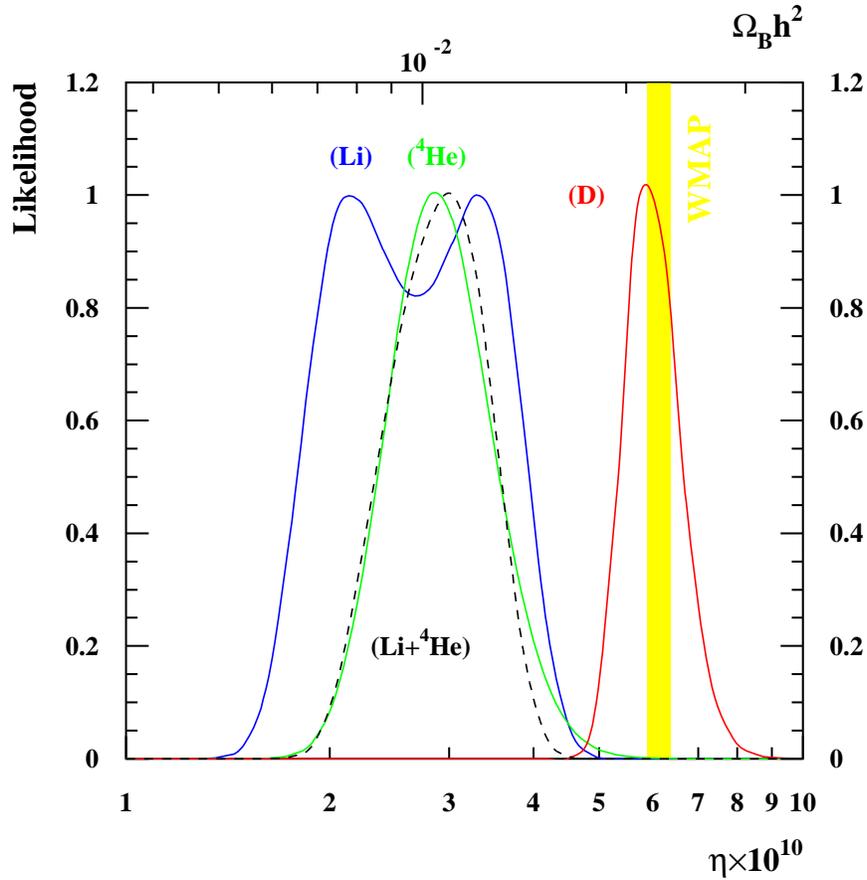}
\caption{Likelihood functions for \deu, \qua\ and \sep\ (solid lines) 
obtained from our SBBN calculations and Kirkman et al.\protect\cite{Kir03},
Izotov and Thuan\protect\cite{Izo04}, and Ryan et 
al.\protect\cite{Rya00} data for \deu, \qua\ and \sep\ respectively.
The dashed curve represent the likelihood function for \qua\ and \sep\ while
the verical stripe shows the WMAP  \obh\ range\protect\cite{WMAP}.}
\label{f:like}
\end{figure}    

Fig.~\ref{f:like} shows a comparison between \obh\ ranges deduced either
from SBBN or WMAP. The curves represent likelihood functions obtained 
from our SBBN calculations and observed deuterium\cite{Kir03}, 
helium\cite{Izo04} and lithium\cite{Rya00} primordial abundances. 
These were obtained as in our previous analysis\cite{Coc02} except for the 
new reaction rates and new \deu\ and \qua\ primordial abundances.
The incompatibility between the \deu\ and \sep\ likelihood curves is more
obvious than before due to the lower D/H adopted value (Kirkman et al., 
averaged value). On the contrary, the new \qua\ adopted value\cite{Izo04}
is perfectly compatible with the \sep\ one as shown on Fig.~\ref{f:like} 
(likelihood curves) and Fig.~\ref{f:bbn} (abundances).
Putting aside, for a moment, the CMB results on the baryonic density, we 
would deduce the following 68\% c.l. intervals:  1.85$<\eta_{10}<$3.90 
[0.007$<$\obh$<$0.014] from \sep\ only 
or 5.4$<\eta_{10}<$6.6 [0.020$<$\obh$<$0.024] from \deu\ only.
If we now consider \qua\ together with \sep, we obtain
2.3$<\eta_{10}<$3.5 [0.009$<$\obh$<$0.013]. 
Hence, including these new \qua\ observations favors a low \obh\ interval 
as proposed in our previous work\cite{Coc02}.
The WMAP result on the contrary definitively favors the upper (\deu) one.
If we now assume that the \qua\ constrain is not so tight, because e.g. of
systematic errors on this isotope whose weak sensitivity to \obh\ requires
high precision abundance determinations, the origin of the discrepancy on 
\sep\ remains a challenging issue very well worth further investigations. 

\section{Possible origins of \sep\ discrepancy between SBBN and CMB}

\subsection{Stellar}

Both observers and experts in stellar atmospheres agree to consider that the 
abundance determination in halo stars, and more particularly that of lithium 
requires a sophisticated analysis. 
The derivation of the lithium abundance in halo stars with
the high precision needed requires a fine knowledge of the physics of stellar 
atmosphere (effective temperature scale, population of different ionization 
states, non LTE (Local Thermodynamic Equilibrium) effects and 1D/3D model 
atmospheres\cite{Asp03}.
However, the 3D, NLTE abundances are very similar to the 1D, LTE results,
but, nevertheless, 3D models are now compulsory to extract lithium abundance
from poor metal halo stars\cite{Bar03}.

Modification of the surface abundance of $Li$ by nuclear burning
all along the stellar evolution has been discussed for a long time in the
literature. 
There is no lack of phenomena to disturb the $Li$ abundance:
rotational induced mixing, mass loss,...\cite{Thea01,Pin02}. 
However, the flatness of the plateau over three decades in metallicity and the 
relatively small dispersion of data represent a real challenge to stellar 
modeling. 
In addition, recent observations of \six\ in halo stars (an even more 
fragile isotope than \sep) 
constrain more severely the potential destruction of lithium\cite{EVF99}.

\subsection{Nuclear}

\begin{figure}[htb]
\includegraphics[width=12cm]{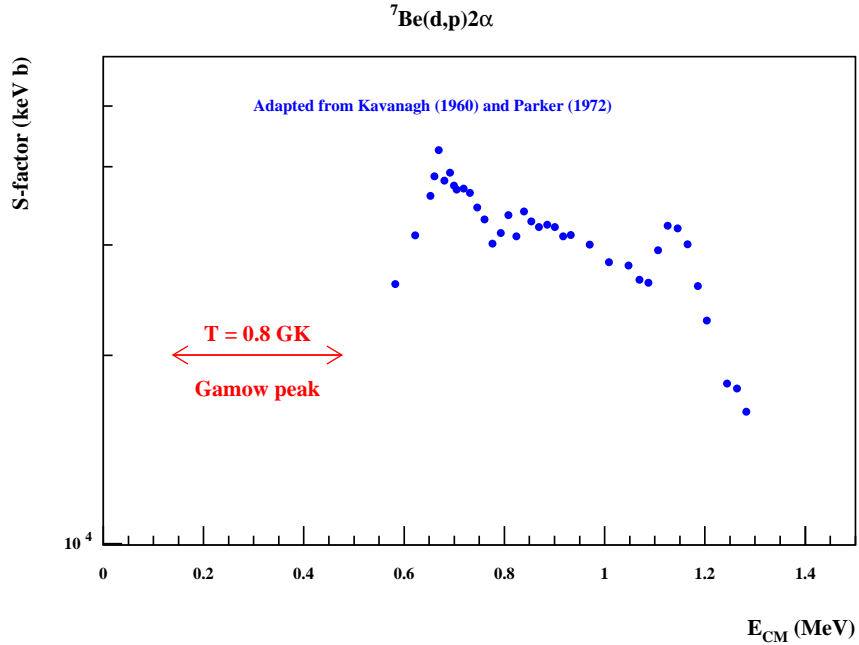}
\caption{The only experimental data available for the $^7$Be(d,p)2$^4$H
reaction from Kavanagh (1960). The displayed $S$--factor is calculated 
as in Parker (1972) from the differential cross section at 90$^\circ$  
($\times4\pi$) leading to the ground and first $^8$Be excited states.  
Note that no data is available at SBBN 
energies as shown by the Gamow peak for a typical temperature of T$_9$ = 0.8
}
\label{f:nucl}
\end{figure}

Large systematic errors on the 12 main nuclear cross sections are 
excluded\cite{Des03,Coc03}.   
However, besides the 12 reactions classically considered in SBBN, 
first of all the influence of {\it all} nuclear reactions needs to be 
evaluated\cite{Coc03}.
It is well known that the valley shaped curve representing $Li/H$ as a 
function of $\eta$ is due to two modes of \sep\ production.
One, at low $\eta$ produces \sep\ directly via $^3$H($\alpha,\gamma)^7$Li
while \sep\ destruction comes from $^7$Li(p,$\alpha)^4$He.
The other one, at high  $\eta$, leads to the formation of $^7$Be 
through $^3$He($\alpha,\gamma)^7$Be while
$^7$Be destruction by $^7$Be(n,p)$^7$Li is inefficient because of the 
lower neutron abundance at high density ($^7Be$ later decays to \sep).
Since the WMAP results point toward the high $\eta$ region,  
a peculiar attention should be paid to $^7$Be synthesis.  
In particular, the $^7$Be+d reactions could be an alternative 
to $^7$Be(n,p)$^7$Li for the destruction of $^7Be$, 
by compensating the scarcity of neutrons at high $\eta$.
Fig.~\ref{f:bbn} shows (dash--dotted lines) that an increase of the 
$^7$Be(d,p)2$^4$He reaction rate by factors of 100 to 300 would 
remove the discrepancy.
The rate for this reaction\cite{CF88} can be traced to an estimate by 
Parker\cite{Par72} who assumed for the astrophysical $S$--factor a 
constant value of 10$^5$~kev.barn. based on the single experimental 
data available\cite{Kav60}.
To derive this $S$--factor, Parker used this measured differential cross 
section at 90$^\circ$ and assumed isotropy of the cross section. 
Since Kavanagh measured only the p$_0$ and p$_1$ protons 
(i.e. feeding the $^8$Be ground and first excited levels), Parker 
introduced an additional but arbitrary factor of 3 to take into account
the possible population of higher lying levels. Indeed, a level at 11.35~MeV
is also reported\cite{Ajz88}. This factor should also include the 
contribution of another open channel in $^7$Be+d: $^7$Be(d,$\alpha)^5$Li
for which no data exist. 
In addition, one should note that {\it no} experimental data
for this reaction is available at energies relevant to $^7$Be Big Bang 
nucleosynthesis (Fig.~\ref{f:nucl}), 
taking place when the temperature has dropped below 10$^9$~K.
A seducing possibility\cite{Coc03} to reconciliate, SBBN, \sep\ and 
CMB observations would then be that new experimental data below $E_d$ = 
700~keV ($E_{cm}\approx$0.5~MeV) 
for $^7$Be(d,p)2$^4$He [and $^7$Be(d,$\alpha)^5$Li]
would lead to a sudden increase in the $S$--factor 
as in $^{10}$B(p,$\alpha)^{7}$Be\cite{Ang93,NACRE}.
This is not supported by known data, but considering 
the cosmological or astrophysical consequences, this is definitely an
issue to be investigated and an experiment is planned in 2004 at 
the Cyclotron  Research  Centre in Louvain-la-Neuve.

\subsection{Cosmology}

Recent theories that could affect BBN include the variation of the fine 
structure constant\cite{Nol02}, the modification of the expansion rate 
during BBN induced by quintessence\cite{Sal02}, modified gravity\cite{Nav02},
or leptons asymmetry\cite{Ori02}. However, their effect is in general 
more significant on \qua\ than on \sep.  

It may not be excluded that some bias exists in the analysis of CMB 
anisotropies. For instance, it has been argued\cite{Gio03} that a 
contamination of CMB map by blazars could affect the second peak of 
the power spectrum on which the CMB \obh\ values are based.

\subsection{Pregalactic evolution}

We note that between the BBN epoch and the birth of the now observed 
halo stars, $\approx$1~Gyr have passed. Primordial abundances could have 
been altered during this period.
For instance, cosmological cosmic rays, assumed to have been 
born in a burst at some high redshift, could have modified these 
primordial abundances in the intergalactic medium\cite{Mon77}. 
This would increase the primordial \sep\ and \deu\ abundances trough 
spallative reactions, increasing in the same time the discrepancy
between SBBN calculations and observations instead to reconciliate them.

Another source of alteration of the primordial abundances could be 
the contribution of the first generation stars (Population III). However, 
it seems difficult that they could reduce the \sep\ abundance without 
affecting the \deu\ one, consistent with CMB \obh.

\section{Conclusions}

The baryonic density of the Universe as determined by the analysis of the 
Cosmic Microwave Background anisotropies is in very good agreement with 
Standard Big--Bang Nucleosynthesis compared to \deu\ primordial abundance
deduced from cosmological cloud observations. However, it strongly 
disagrees with lithium observations in halo stars (Spite plateau)
and possibly \qua\ new observations.
The origin of this discrepancy, if not nuclear, is a challenging issue.

\end{document}